\documentclass[intlimits,twoside,a4paper]{article}

\usepackage[utf8]{inputenc}
\usepackage[eqsecnum]{cmpj3}

\usepackage{bm}

\issue{2025}{28}{3}{33703}
\doinumber{10.5488/CMP.28.33703}

\title[Tripartite entanglement in mixed-spin triangle trimmer]%
{Tripartite entanglement in mixed-spin triangle trimmer}
\author[Zh. Adamyan, V. Ohanyan]{Zh. Adamyan\orcid{0000-0001-5937-2675}, V. Ohanyan\orcid{0000-0002-7810-7321}}
\address{
 Laboratory of Theoretical Physics, Yerevan State University, 1 Alex Manoogian St., 0025 Yerevan, Armenia;
 CANDLE, Synchrotron Research Institute, 31 Acharyan St., 0040 Yerevan, Armenia
}

\date{Received July 2, 2025, in final form August 12, 2025}
\begin{document}

\maketitle

\begin{abstract}
Heisenberg model spin systems offer favourable and manageable physical settings for
generating and manipulating entangled quantum states. In this work mixed
spin-(1/2,1/2,1) Heisenberg spin trimmer with two different but isotropic Land\'e $g$-factors and two
different exchange constants is considered. The study undertakes the task of finding the optimal parameters to
create entangled states and control them by external magnetic field. The primary objective of this work is to examine the tripartite entanglement of a system and
the dependence of the tripartite entanglement on various system parameters. Particularly, the effects of non-conserving
magnetization are in the focus of our research. The source of non-commutativity between the magnetic moment operator and the Hamiltonian is non-uniformity of $g$-factors. To quantify the tripartite entanglement, an entanglement measure
refered to as ``tripartite negativity'' has been used in this work.
%
%
\printkeywords
\end{abstract}

\section{Introduction}

Quantum entanglement, as a fundamental phenomenon in quantum mechanics, has recently attracted increasing attention  due to its key role in quantum communication and information processing paradigms~\cite{ent,guhne,bephys54,bephys76,ekert}. This non-classical correlation serves as a cornerstone for various quantum technologies, including quantum teleportation \cite{bephys70, bouwlondon, chophys95, jinnature, baurphys, bjphysrev}, quantum computing \cite{kokphysmod, kimphysrev, brodphysrev}, and quantum cryptography~\cite{vedral89, avellaphysrev}. Moreover, the study of entanglement has yielded significant insights in diverse fields, ranging from black hole physics, where it has facilitated progress in applying quantum field theory methods, to the investigation of quantum phase transitions and collective phenomena in many-body systems and condensed matter physics.
In the last decades, significant research efforts have focused on investigating the entanglement properties of quantum spin clusters and molecular magnets \cite{molmag, kahn}, motivated by compelling evidence suggesting that molecular magnets could serve as promising candidates for a physical realization of qubits in quantum information technologies. At the moment, several schemes of quantum computers working with molecular magnets are proposed \cite{los01, ste08, ses15}. This rapidly growing field has generated a substantial body of literature exploring various aspects of quantum entanglement in many-body spin systems \cite{ana11, ana12, str14, car18, sou19, sou20, ada20, cen20, eki20, kar20, str20, gal21, gal21b, gal22, ben22, zad22, zhe22, mer, ghan24}, contributing to our understanding of quantum correlations at the molecular scale and their potential applications in quantum information science.

Magneto-thermal properties of single molecule magnets (SMM) are particularly sensitive to the situation when magnetic moment operator does not commute with the Hamiltonian, that gives rise to a non-conserving magnetization.  The most common reason for the non-conserving magnetization is the different $g$-factors of different magnetic ions within the molecule \cite{sou19, ada20, cen20, str05, vis09, van10, bel14, oha15, tor16, tor18, var19, kro20, pan20, jap21, bar23}. Non-conserved magnetization can affect the magnetization curve drastically. When the magnetic moment
is a good quantum number (conserving magnetization operator), then for the SMM, the
magnetization curve at zero temperature consists of a series of horizontal parts
(magnetization plateaus) with step-like transitions between them. Each plateau
corresponds to a certain eigenstate which is the ground state at given values of
magnetic field. A constant value of the magnetic moment at each plateau is the expectation value of the magnetization operator for a given eigenstate. Transitions between plateaus correspond to level-crossing points.
However, if the magnetization operator does not commute with the Hamiltonian,
the magnetic field dependence within the given ground state can be continuous, since eigenstates with the given value of energy are not simultaneously eigenstates of the magnetic moment operator. Thus, even at zero temperature, the magnetization curve of SMM with non-conserving magnetization has much in common with magnetization curve of a real
many-body system \cite{bel14, oha15, var22}. Non-uniform $g$-factors can bring to drastic change of the eigenstates of finite spin cluster in comparison with the eigenstates of the same Hamiltonian but with uniform $g$-factors. These changes lead to incoherent superpositions of the spin states basic vectors with coefficients dependent on the 
magnetic field magnitude and other parameters of the system. The latter, in its turn, opens up new
possibilities for manipulating the quantum and/or thermal entanglement by means of magnetic field.

This study focuses on a mixed spin-($1/2,1/2,1$) trimmer system with two distinct, yet isotropic, exchange interaction  constants and non-conserved magnetization due to non-uniform $g$-factors. In our previous work we  examined the properties of a bipartite entanglement for this system \cite{mer,zh}. In this work we are going to investigate the tripartite entanglement of the system with the aid of the tripartite negativity measure. We analyse the dependence of negativity on various system parameters, including exchange interaction  constants, non-conserved magnetization, and external magnetic field. A comparative analysis is conducted between the scenario involving non-conserved magnetization and the case with homogeneous $g$-factors.
The investigation encompasses several exchange interaction  constants configurations:

A ferromagnetic interaction case ($J_1<0$, $J_2<0$),
a scenario with two antiferromagnetic interactions $(J_1=J_2>0$, $J_2>J_1>0)$ and
two mixed cases combining ferromagnetic and antiferromagnetic interactions($J_1>0$, $J_2<0$ and $J_1<0$, $J_2>0$).

The paper is organized as follows. In the second section we introduce the quantum spin model and present its exact spectrum and eigenstates. The next third section is devoted to tripartite negativity. In  section~\ref{sec4} we calculate tripartite negativity and present the plots of its magnetic field behaviour. The paper ends with conclusion.

\section{The model}
We consider a mixed spin trimmer with spins 1/2, 1/2, and 1 in a triangular arrangement, with two different exchange couplings $J_{1}$ (between spin-1 and each spin-1/2) and $J_2$ (between spin-1/2 pairs) and two different $g$-factors, the one of 1/2 spins and the spin-1 ion have $g$-factor equal to $g_1$, while the other spin-1/2 ion  has $g$-factor equal to $g_2$.
\begin{figure}[!t]
\begin{center}
 \includegraphics[width=70mm]{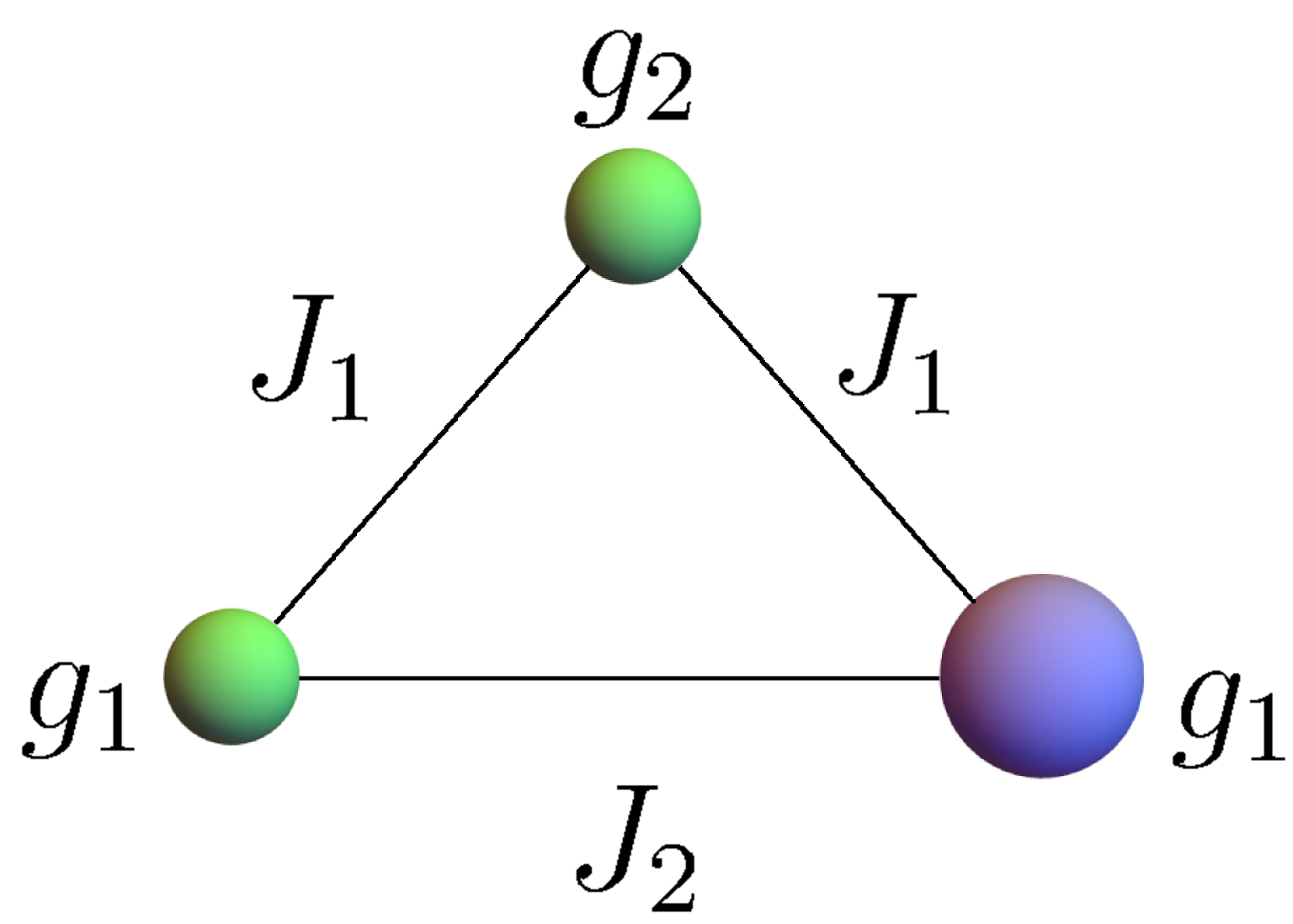}
  \caption{(Colour online) Symbolic picture of triangular trinuclear mixed-spin magnetic molecule with spins 1/2, 1/2 (small balls), and 1 (large ball). The Lang\`{e} $g$-factors for one spin-1/2 ion and spin-1 ion are supposed to be the same ($g_1$), whereas the $g$-factor for the second spin-1/2  differs from them, $g_2$.}
    \label{fig1}
  \end{center}
\end{figure}
The spin Hamiltonian of the model has the following form:
\begin{equation}\label{ham}
\mathcal{H} = J_{1}\left(\boldsymbol s_{1}\boldsymbol s_{2}+\boldsymbol s_{2}\boldsymbol S_{3}\right)+J_{2}\boldsymbol s_{1}\boldsymbol S_{3}-B\left(g_{1}s_{1}^z+g_{2}s_{2}^z+g_{1}S_{3}^z\right),
\end{equation}
where $\boldsymbol s_{a}$, $a=1, 2$, are spin-1/2 operators, and  $\boldsymbol S_{3}$ stands for spin-1 operators.

 An unusual choice of the interaction constants and $g$-factors distribution should be explained. On the one hand, we aimed to have a model that admits analytical solution in a possible simple way. The arrangement of the system parameters in the Hamiltonian (\ref{ham})
admits two permutation symmetry operations, $P_{12}$ and~$P_{13}$. Owing
to these symmetries, the eigenvalue problem for the Hamiltonian can be
factorized, allowing an analytical solution in terms of only quadratic
equations. Remarkably, this is one of the very few possible Hamiltonians
for mixed-spin $(\tfrac{1}{2}, \tfrac{1}{2}, 1)$ triangular clusters
that permits a fully analytical treatment of the eigenvalues. On the other hand, the model chosen in the present research can have a physical realization in optical lattices with trapped fermionic atom (ions) \cite{boh} which offers a broad opportunity to construct the effective models of quantum magnets and strongly correlated electrons.  The principal advantage of this technology is the possibility of manipulating the parameters of exchange interaction using laser beam properties. In particular, optical lattices offer an exceptionally accurate
realization of the Hubbard model, which, at half-filling and in the
regime of strong on-site repulsion, can be mapped onto the Heisenberg
model with the standard effective exchange coupling $J_{\mathrm{eff}} =
{4t^2}/{U}$,
where $t$ is the tunneling amplitude, determined by the relative height
and width of the laser-field barriers between trapped fermionic atoms,
and $U$ is the on-site repulsion energy. The latter can be tuned via a
Feshbach resonance \cite{boh}.  This setup is particularly relevant to our present results, since atoms in optical lattices are very well known as feasible examples of quantum bits.

 The eigenvalues and the eigenstates of the Hamiltonian are as follows \cite{mer}:
\begin{eqnarray}
{E}_{1,2}&=&\frac{1}{4}\left[3J_{1}+2J_{2}\mp2B(3g_{1}+g_{2})\right], \quad
{E}_{3,4}=\frac{1}{4}\left[J_{1}-4J_{2}\mp2B(g_{1}+g_{2})\right], \nonumber\\
{E}_{5,6}&=&\frac{1}{4}\left(-J_{1}-4J_{2}\mp2\sqrt{B^{2}g_{-}^2+J_1^2 }\right),\quad
{E}_{7,8}=\frac{1}{4}\left(-J_{1}+2J_{2}\mp2\sqrt{B^{2}g_{-}^2+4J_1^2}\right),\nonumber \\
{E}_{9,10}&=&\frac{1}{4}\left(-J_{1}+2J_{2}-4Bg_{1}\mp2Q^+\right),\quad
{E}_{11,12}=\frac{1}{4}\left(-J_{1}+2J_{2}+4Bg_{1}\mp2Q^-\right), 
\label{eigval}
\end{eqnarray}
where $Q^\pm = \sqrt{(Bg_{-}\pm J_{1})^2+3J_{1}^2}$, $\quad g_{-}=g_1-g_2$. 

\begin{eqnarray}
&& |\psi_{1,2}\rangle= \left|\pm\frac{1}{2},\pm\frac{1}{2},\pm1\right\rangle,  \nonumber\\
 && |\psi_{ 3,4}\rangle=\frac{1}{\sqrt{3}}
\left(\pm\sqrt{2}\left|\mp\frac{1}{2},\pm\frac{1}{2},\pm1\right\rangle \mp\left|\pm\frac{1}{2},\pm\frac{1}{2},0\right\rangle\right),  \nonumber \\
&&  |\psi_{ 5,6}\rangle= \frac{1}{\sqrt{3(1+M^{2}_{\pm})}}
\left[M^{\pm}\left(\sqrt{2}\left|\frac{1}{2},\frac{1}{2},-1\right\rangle- \left|-\frac{1}{2},\frac{1}{2},0\right\rangle\right)+\sqrt{2}\left|-\frac{1}{2},-\frac{1}{2},1\right\rangle- \left|\frac{1}{2},-\frac{1}{2},0\right\rangle\right], \nonumber \\
&&|\psi_{ 7,8}\rangle= \frac{1}{\sqrt{3(4+K^{2}_{\mp})}}
\left[K^{\mp}\left(\left|\frac{1}{2},\frac{1}{2},-1\right\rangle+\sqrt{2}\left|-\frac{1}{2},\frac{1}{2},0\right\rangle\right)+
2\left|-\frac{1}{2},-\frac{1}{2},1\right\rangle+ 2\sqrt{2}\left|\frac{1}{2},-\frac{1}{2},0\right\rangle\right], \nonumber \\
  &&|\psi_{ 9,10}\rangle= \frac{1}{\sqrt{3+G^{2}_{\pm}}}
\left(\sqrt{2}\left|\frac{1}{2},\frac{1}{2},0\right\rangle-
G^{\pm}\left|\frac{1}{2},-\frac{1}{2},1\right\rangle+ \left|-\frac{1}{2},\frac{1}{2},1\right\rangle\right), \nonumber \\
&&|\psi_{ 11,12}\rangle= \frac{1}{\sqrt{3+U^{2}_{\mp}}}
\left(\left|\frac{1}{2},-\frac{1}{2},-1\right\rangle+
U^{\mp}\left|-\frac{1}{2},\frac{1}{2},-1\right\rangle+ \sqrt{2}\left|-\frac{1}{2},-\frac{1}{2},0\right\rangle\right),\nonumber\\
&&M^\pm=\frac{-Bg_{-}\pm\sqrt{B^2g_{-}^2+J_{1}^2}}{J_{1}},\nonumber \quad
K^\pm=\frac{Bg_{-}\pm\sqrt{B^2g_{-}^2+4J_{1}^2}}{2J_{1}}, \nonumber \\
&&G^\pm=\frac{Bg_{-}+J_{1}\pm Q^+}{J_{1}},
\quad U^\pm=\frac{Bg_{-}-J_{1}\pm Q^-}{J_{1}}, 
\label{eq2.3}
\end{eqnarray}
where the standard Ising basis is chosen, $\left|s_1^z, s_2^z,
S_3^z\right\rangle$. Since one of the main parts of this research is devoted to the comparison of entanglement properties of the system with non-conserving magnetization and its uniform-$g$ counterpart, we have to use the corresponding eigenvectors for calculating the entanglement measures. However, the eigenvectors given above not always admit a continuous limit at $g_2\rightarrow g_1$. For the case of uniform $g$-factors, the corresponding Hamiltonian should be diagonalized separately.  For the case $J_1=J_2$ and $g_2\rightarrow g_1$, the eigenvectors  $|\psi_{3,4}\rangle$, $|\psi_6\rangle$, $|\psi_7\rangle$ $|\psi_9\rangle$ and $|\psi_{11}\rangle$ change noncontinuously. They acquire the following form:
\begin{eqnarray}
&&|\psi_{3,4}\rangle_0=\frac 12 \left(\left|\frac 12, -\frac 12, \pm 1\right\rangle+\left|-\frac 12, \frac 12, \pm 1\right\rangle-\sqrt 2\left|\pm\frac 12, \pm\frac 12, 0\right\rangle\right),\nonumber\\
&& |\psi_ 6\rangle_0=\frac{1}{\sqrt{2}}\left( \left|-\frac{1}{2},-\frac{1}{2},1\right\rangle-\left|\frac{1}{2},\frac{1}{2},-1\right\rangle \right), \nonumber  \:\:\:  |\psi_7\rangle_0=\frac{1}{\sqrt{2}}\left(\left|-\frac{1}{2},\frac{1}{2},0\right\rangle-\left|\frac{1}{2},-\frac{1}{2},0\right\rangle \right), \nonumber \\
&&|\psi_9\rangle_0=\frac{1}{\sqrt{2}} \left(\left|-\frac 12, \frac 12, 1\right\rangle-\left|\frac 12, -\frac 12, 1\right\rangle\right), \quad|\psi_{11}\rangle_0=\frac{1}{\sqrt{2}} \left(\left|-\frac 12, \frac 12, -1\right\rangle-\left|\frac 12, -\frac 12, -1\right\rangle\right). 
\label{psi0}
\end{eqnarray}
The rest of the eigenvectors change continuously under $g_2\rightarrow g_1$. The origin of this phenomenon lies in the symmetry restoration, which
takes place under a uniform distribution of the exchange coupling and
$g$-factors. Additional $P_{23}$ permutation symmetry element appears when
$J_1=J_2$ and $g_1=g_2$.
A detailed examination of system phase diagrams and magnetic properties was given in our previous work \cite{mer}.

\section{Tripartite negativity}

Several different entanglement measures can be used to quantify the quantum entanglement~\cite{ent, horodecki}. However, for the mixed-spin clusters, negativity \cite{vidal} is the most convenient measure, as it can be easily constructed and calculated for any pair of spins with the aid of reduced density matrix. For the system under consideration, three pairwise bipartite negativities can be constructed. The numerical value of negativity, which varies from 0 (no entanglement) to $1/2$ (maximally entangled pair), corresponding to the $i$-th and $j$-th particles of the system,  $\mathit{Ne}_{ij}$, is equal to the sum of absolute values of negative eigenvalues of partially transposed reduced two-particle density matrix, $\rho_{ij}^T$, which is constructed in the following way:
\begin{eqnarray}\label{neg1}
&&\left\langle \tilde{\xi_i}, \xi_j \right| \rho_{ij}^T \left| \xi_i, \tilde{\xi_j} \right\rangle= \left\langle \xi_i, \xi_j  \left| \rho_{ij} \right| \tilde{ \xi_i},
\tilde{\xi_j}\right\rangle, \\
&&\rho_{ij}=\sum_{\xi_k}\left\langle \xi_k \left| \rho \right|   \xi_k \right\rangle,\quad k\neq i,j ,\nonumber
\end{eqnarray}
where $\left|\xi_i, \xi_j, \xi_k \right\rangle$ is a standard basis for the states of $\boldsymbol{s}_1, \boldsymbol{s}_2$ ($\xi_i=\pm1/2$) and $\boldsymbol{S}_3$ ($\xi_i=-1, 0, 1$) spins. Then, the negativity is obtained according to
\begin{equation}\label{neg2}
\mathit{Ne}_{ij}=\sum\limits_{a}|\mu_{a}|.
\end{equation}		
Bipartite negativity accounts for the entanglement between pairs of subsystems within the given system. In order to clarify the overall entanglement of three subsystems one can use the so-called tripartite negativity~\cite{trip}. In general, in the system of three particles, three different distributions of the entanglement can be found among its configurations: fully separable state, exhibiting no entanglement, three states where the pair of particles are entangled, but the third particle is not (biseparable states) and one configuration in which all three particles are in the entangled state (tripartite entanglement) \cite{trip}. Tripartite negativity serves as an appropriate measure for characterizing the numerical degree of tripartite entanglement. This measure is defined by the following expression:
\begin{equation}
\mathit{Ne}_{ABC}=(\mathit{Ne}_{A-BC}\,\mathit{Ne}_{B-AC}\,\mathit{Ne}_{C-AB})^{1/3},
\label{ABC}
\end{equation}
where $\mathit{Ne}_{A-BC}$, $\mathit{Ne}_{B-AC}$, $\mathit{Ne}_{C-AB}$ are generalized bipartite negativities between the corresponding spin and the rest of the system. This quantity is calculated according to the similar formulas as given in equations~(\ref{neg1})--(\ref{neg2}), when the trace in the equation (\ref{neg1}) is not taken,
\begin{eqnarray}
&&\left\langle \tilde{\xi_i}, \xi_j, \xi_k \right| \rho_{i-jk}^T \left| \xi_i, \tilde{\xi_j}, \tilde{\xi_k} \right\rangle= \left\langle \xi_i, \xi_j, \xi_k  \left| \rho_{ijk} \right| \tilde{ \xi_i},\tilde{\xi_j}, \tilde{\xi_k}\right\rangle.
\end{eqnarray}
The multiplicative tripartite negativity serves as both a necessary and
sufficient criterion for genuine tripartite entanglement in pure states.
However, for mixed states in systems with dimensions exceeding $2\times3$ there exist entangled states with zero negativity.
Consequently, while the tripartite negativity ($\mathit{Ne}_{ABC}$) remains a
necessary condition for detecting entanglement in mixed states, it is
not sufficient for complete entanglement characterization. Importantly,
$\mathit{Ne}_{ABC}>0$ constitutes a sufficient condition for
Greenberger–Horne–Zeilinger (GHZ)-distillability, the capability to
asymptotically distill genuine multipartite entanglement into GHZ
states. This feature is of particular relevance to quantum computation.
Therefore, despite its incapability to fully resolve the separability
problem, tripartite negativity remains a valuable tool for
characterizing entanglement in mixed states, especially for identifying
operationally useful, distillable entanglement resources.

For mixed spin trimer (1/2, 1/2, 1), tripartite negativity can take values from 0 (no entanglement) to $\sqrt[3]{\frac 14}\approx 0.63$ (maximally entangled state).
Here, we deal only with purely quantum or zero-temperature entanglement, so the density matrix, $\rho$, we are working with is defined for each of the twelve eigenstates of the Hamiltonian as a pure-state density matrix:
\begin{equation}
\rho_{i}=|\Psi_{i}\rangle \langle \Psi_{i}|, \;\; i=1,...,12.
\end{equation}
In case of $n$ degenerate eigenstates, one should use
\begin{eqnarray}
\rho_{i_1...i_n}=\frac {1}{n}\sum_{a=1}^n|\Psi_{i_a}\rangle \langle \Psi_{i_a}| .
\end{eqnarray}

\section{Results}\label{sec4}
According to the definition given above [equation (\ref{ABC})] we have obtained analytic expressions for the tripartite negativity for the eingenstates of the system under consideration. We present here only those which are relevant for the ground states phase diagram, obtained in our previous work \cite{mer}. Since tripartite negativity is given by the equation (\ref{ABC}), it is sufficient to know only all generalized bipartite negativities, $Ne^i_{A_{i} - A_{j}A_{k}}$. The index $i$ stands for the eigenstate numbering the parameter. Bellow, we denote spin-1/2 ion with $g$-factor equal to $g_1$ by $A$, spin-1/2 ion with $g$-factor equal to $g_2$ by $B$ and ion with spin-1 by $C$.  Interestingly, some eigenstates can have the same $Ne^i_{A_{i} - A_{j}A_{k}}$.
\begin{figure}[h]
	\begin{center}
		\includegraphics[width=120mm]{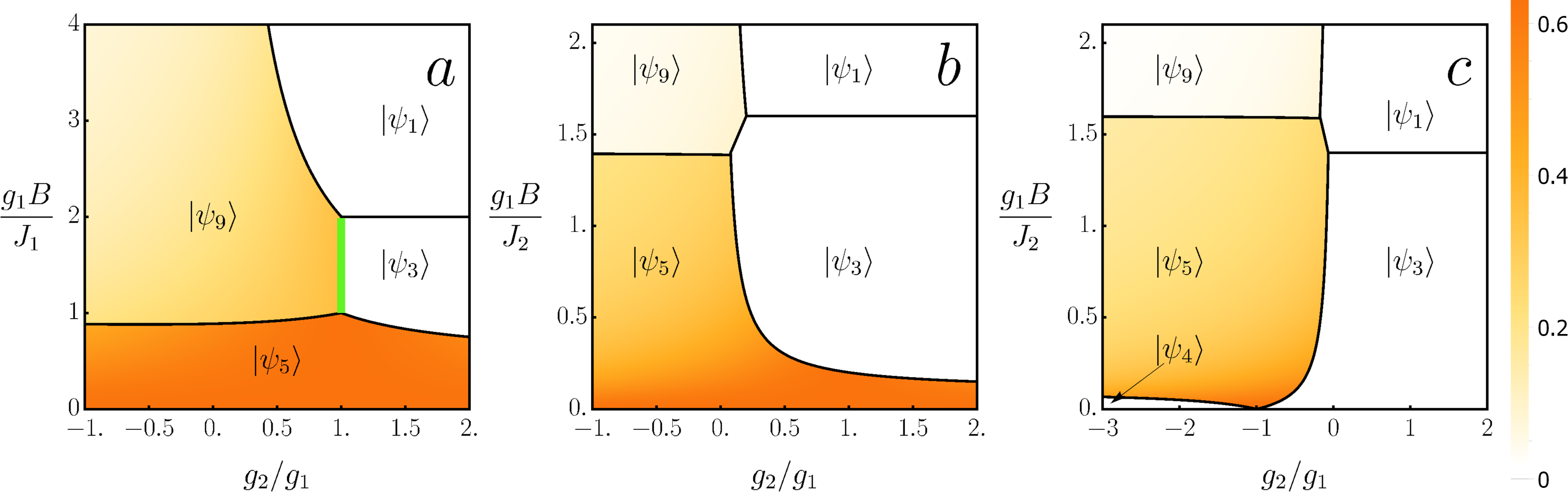}
		\caption{(Colour online) Density plots of tripartite negativity for $J_1=J_2>0$ (panel a),  $J_2>0\,\, \text{and}\,\, J_1=\frac{1}{5}J_2$ (panel b) and $ J_2>0\,\, \text{and}\,\, J_1=-\frac{1}{5}J_2$ (panel c) cases.}
		\label{fig2}
	\end{center}
\end{figure}
\begin{figure}[h]
	\begin{center}
		\includegraphics[width=90mm]{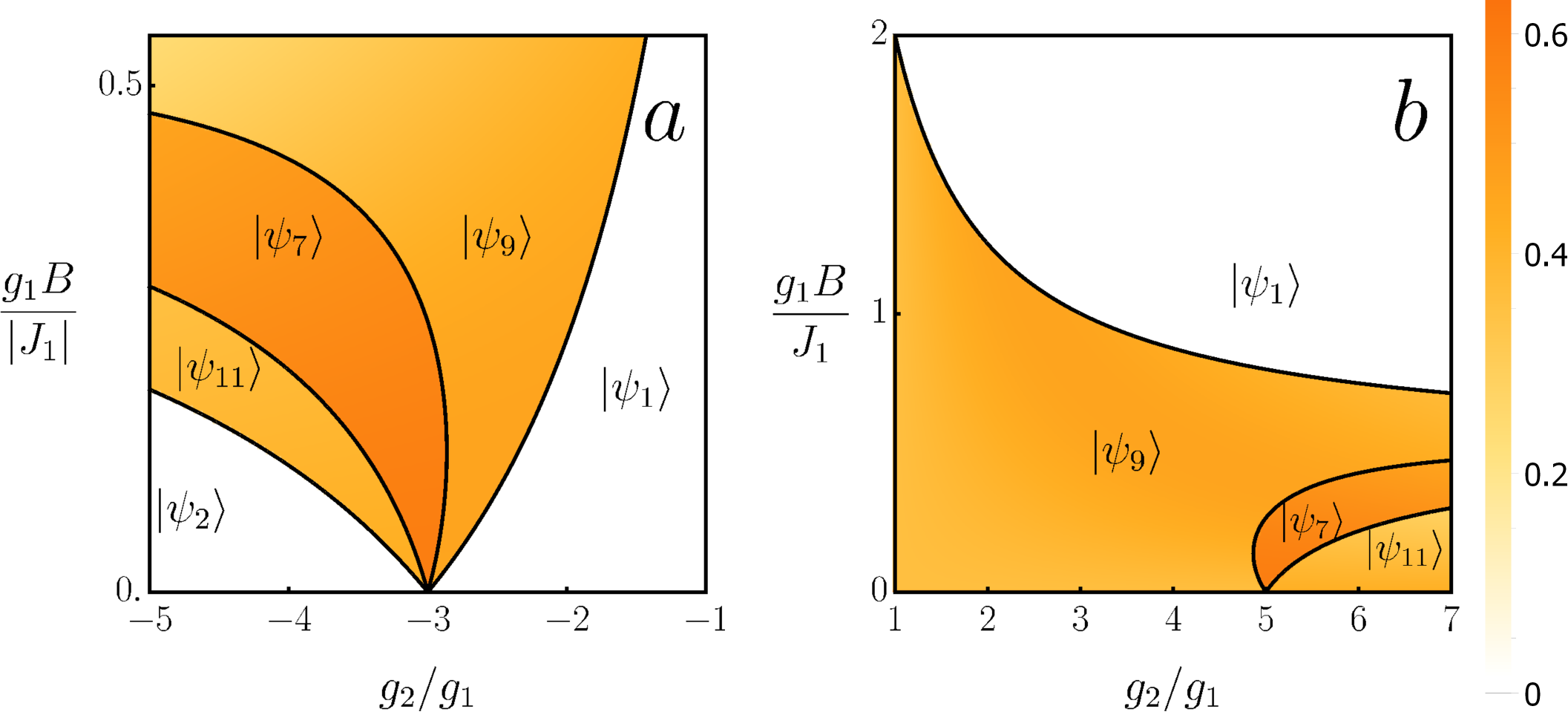}
		\caption{(Colour online) Density plots of tripartite negativity for $J_1<0$, arbitrary $J_2\leqslant 0$ and $J_1>0$, arbitrary $J_2<0$ cases in panels a and b, respectively.}
		\label{fig3}
	\end{center}
\end{figure}
	\allowdisplaybreaks
\begin{eqnarray}
&& Ne^{3,4}_{A - BC} = Ne^{3,4}_{C - AB} = \frac {\sqrt {2}} {3},
 \:\:\: \:\:\:Ne^{3,4}_{B - AC} = 0, \:\:\: \:\:\:Ne^{5,6}_{A - BC} = \frac {\sqrt {2 M_{\pm} ^4 + 5 M_{\pm} ^2 + 2}} {3\left (M_{\pm} ^2 +
      1 \right)},\nonumber\\
&&Ne^{5,6}_ {B - AC} = \left | \frac {M_{\pm}} {M_{\pm} ^2 + 1} \right  |,  \:\:\: \:\:\: Ne^{5,6}_ {C - AB} = \frac {2} {3}\left | \frac {M_{\pm}} {M_{\pm} ^2 +
      1} \right  | +\frac {\sqrt {2}\sqrt {M_{\pm} ^4 +
         M_{\pm} ^2}} {3\left (M_{\pm} ^2 +
         1 \right)} + \frac {\sqrt {2}} {3\sqrt {M_{\pm} ^2 + 1}}, \nonumber\\
&&Ne^{7,8}_ {A - BC} = \frac {\sqrt {2}\sqrt {K_{\mp} ^4 + 10 K_{\mp} ^2 +
      16}} {3\left (K_{\mp} ^2 + 4 \right)}, \:\:\: \:\:\:Ne^{7,8}_ {B - AC} = 2\left | \frac {K_{\mp}} {K_{\mp} ^2 + 4} \right |, \nonumber\\
 &&  Ne^{7,8}_ {C - AB} = \frac {2} {3}\left | \frac {K_{\mp}} {K_{\mp} ^2 +
      4} \right | +\frac {\sqrt {2}\sqrt {K_{\mp} ^4 +
         4 K_{\mp} ^2}} {3\left (K_{\mp} ^2 +
         4 \right)} + \frac {2\sqrt {2}} {3\sqrt {K_{\mp} ^2 + 4}}, \:\:\: \:\:\:Ne^{9,10}_ {A - BC} = \frac {\sqrt {G_{\pm} ^2 + 2}} {G_{\pm} ^2 + 3}, \nonumber\\
&&Ne^{9,10}_{B - AC} = \sqrt {3}\left | \frac {G_{\pm}} {G_{\pm} ^2 +
      3} \right  |, \:\:\: \:\:\: Ne^{9,10}_{C - AB} = \frac {\sqrt {2}\sqrt {G_{\pm} ^2 + 1}} {G_{\pm} ^2 + 3} , \:\:\: \:\:\:
Ne^{11,12}_{A - BC} = \sqrt {3}\left | \frac {U_{\mp}} {U_{\mp}^2 + 3} \right  |, \nonumber\\
  && Ne^{11,12}_ {B - AC} = \frac {\sqrt {U_{\mp}^2 + 2}} {U_{\mp}^2 + 3} , \:\:\: \:\:\:Ne^{11,12}_ {C - AB} = \frac {\sqrt {2}\sqrt {U_{\mp}^2 + 1}} {U_{\mp}^2 + 3}. 
  \label{eq4.1}
\end{eqnarray}
In our previous work, a detailed analysis of the ground states phase diagrams of the system were presented~\cite{mer}. Here, we demonstrate only those of them, which exhibit interesting features of the tripartite negativity dependent on the value on non-uniform $g$-factor. The main purpose of the present research is to figure out how non-uniform $g$-factors affect tripartite entanglement properties of the model. It is worth mentioning that the most remarkable enhancement of the bipartite negativity caused by non-uniform $g$-factors reported in our previous work \cite{mer} concerned the case $J_2=J_1$ and gave almost 7-fold robust increase of $\mathit{Ne}_{13}$ for arbitraryly small difference between $g_2$ and $g_1$. For the tripartite entanglement, the situation is quite different. Here, we presented density plots of the tripartite negativity, $\mathit{Ne}_{ABC}$, projected onto ground state phase diagrams in the ``$g$-factors ration'' -- ``dimensionless magnetic field'' plane.
Two antiferromagnetic cases, $J_1=J_2>0$, $J_2>0\,\, \text{and}\,\, J_1=\frac{1}{5}J_2$,  and a mixed case, $ J_2>0\,\, \text{and}\,\, J_1=-\frac{1}{5}J_2$, are presented in the figure~\ref{fig2}.
 In the case of equal antiferromagnetic coupling (panel a) the phase diagram includes four eigenstates $|\psi_{1}\rangle, |\psi_{3}\rangle,|\psi_{5}\rangle$ and $|\psi_{9}\rangle$. However, two of them, $|\psi_3\rangle$ and $|\psi_9\rangle$, transform non-continuously under
  $g_2\rightarrow g_1$. For uniform situation, $g_2=g_1$, degeneracy line between $|\psi_3\rangle$ and $|\psi_9\rangle$ regions corresponds to the degenerate superposition of $|\psi_3\rangle_0$ and $|\psi_9\rangle_0$ given in the equation (\ref{psi0}). This segment is highlighted in green in the right-hand panel in figure~\ref{fig2}. The value of the tripartite negativity here is $\mathit{Ne}_{ABC}^{(3+9)_0}=\frac 12\sqrt[3]{\frac 34}\approx0.45$.
Interestingly, the $|\psi_{5}\rangle$ eigenstate, which is the zero-field ground state for arbitrary value of $g_2/g_1$,  exhibits maximal three-particle entanglement at $B=0$ and at the segment $g_2=g_1$, $\mathit{Ne}_{ABC}^5=\sqrt[3]{\frac 14}\approx0.63$. The most dramatic discrepancy between bipartite and tripartite entanglement properties  occurrs for the $|\psi_3\rangle$ eigenstate, for which one of the bipartite negativities becomes almost seven times larger than the corresponding $g$-uniform value \cite{mer}. Tripartite negativity is zero for $|\psi_3\rangle$, which means that this state is biseparable. Thus, in the case of $g_2>g_1$, when magnetization is not conserved, it is possible to create two considerably different entanglement regimes and control them by an external magnetic field. The first regime exhibits maximum value for tripartite entanglement, while the second regime exhibits maximum value for bipartite entanglement and has no tripartite entanglement. As usual, for a large enough magnetic field, the system reaches its saturated ($|\psi_1\rangle$) or quasi-saturated ($|\psi_9\rangle$) states with zero or vanishing entanglement. For the case $J_2>0,\,\, J_1=\frac{1}{5}J_2$ (figure~\ref{fig2}, panel b) the system exhibits a similar behavior. There are eigenstates with strong bipartite entanglement, but with zero $\mathit{Ne}_{ABC}$, $|\psi_3\rangle$ and the region where both quantities are quite large, $|\psi_5\rangle$. However, the presence of non-conserving magnetization does not bring any quantitative or qualitative benefits. The only difference between $g_2=g_1$ and $g_2\ne g_1$ cases are that for the second case tripartite negativity is dependent on the magnetic field within the same ground state. Figure~\ref{fig2} (panel c) shows the density plot of tripartite negativity for the $ J_2>0,\,\, J_1=-\frac{1}{5}J_2$ case. Here,  the system has one additional region of ground state corresponding to $|\psi_{4}\rangle$. Here, in contrast to the previous cases, the regime with maximal or essentially large $\mathit{Ne}_{ABC}$ is absent. The system exhibits maximal entanglement under the conditions $g_2=-g_1$ and when magnetic field is close to zero. In the figure~\ref{fig3} the cases $J_1<0$, arbitrary $J_2\leqslant 0$  and $J_1>0$, arbitrary $J_2<0$ are presented. For both cases, maximal possible tripartite negativity is achieved for the eigenstate $|\psi_{7}\rangle$ with value 0.59.

\section{Conclusion}
In the paper we considered the mixed spin (1/2, 1/2, 1) Heisenberg spin trimmer with non-conserving magnetization. Analytical results for tripartite negativity were obtained for all ground states. In general, depending on the relations between two exchange constants, the system can exhibit five regimes of magnetic behavior \cite{mer}. These regimes, in their turn, are storngly affected by the ration of the $g$-factors, $g_2/g_1$. It was shown that in fully antiferromagnetic cases, it is possible to obtain a three-regime system (fully separable, biseparable, tripartite entangled) and control it by magnetic field.  In the case  $ J_2=J_1>0$, these regimes are possible only when $g_2>g_1$, when magnetisation is non-conserving.

\section*{Acknowledgements}
 The authors acknowledge partial financial support form ANSEF (Grants No. PS-condmatth-2884 and PS-condmatth-3273 ) and from CS RA MESCS (Grants No. 21AG-1C047, 21AG-1C006 and 23AA-1C032).


\ukrainianpart

\title{Тричастинкова заплутаність в трикутному тримері зі змішаним спіном}
\author{Ж. Адамян, В. Оганян}
\address{
	Лабораторія теоретичної фізики, Єреванський державний університет, вул. Алекса Манугяна, 1, 0025 Єреван, Вірменія;\\
	CANDLE, Інститут синхротронних досліджень, вул. Ачаряна, 31, 0040 Єреван, Вірменія
}
%
%
%

\makeukrtitle

\begin{abstract}
	\tolerance=3000%
	Спінові системи моделі Гейзенберга пропонують сприятливі та керовані фізичні умови для генерації та маніпулювання заплутаними квантовими станами. У цій роботі розглядається змішаний спін-(1/2,1/2,1) тример спіну Гейзенберга з двома різними, але ізотропними $g$-факторами Ланде та двома різними константами обміну. Дослідження спрямоване на пошук оптимальних параметрів для створення заплутаних станів та керування ними зовнішнім магнітним полем. Основною метою цієї роботи є дослідження тричастинкової заплутаності системи та її залежності від різних параметрів системи. Зокрема, в центрі нашого дослідження знаходяться ефекти намагніченості, яка не зберігається. Джерелом некомутативності між оператором магнітного моменту та гамільтоніаном є неоднорідність $g$-факторів. Для кількісної оцінки тричастинкової заплутаності в цій роботі було використано міру заплутаності, яка називається ``тричастковою негативністю''.
	\keywords тричастинкова заплутаність, одномолекулярні магніти, намагніченість, яка не зберігається
	
\end{abstract}      
      
\lastpage
\end{document}